\documentclass[conference]{IEEEtran}
\IEEEoverridecommandlockouts
\usepackage{cite}
\usepackage{amsmath,amssymb,amsfonts}
\usepackage{algorithmic}
\usepackage{float}
\usepackage{graphicx}
\usepackage{textcomp}
\usepackage{xcolor}
\usepackage[utf8]{inputenc}
\usepackage{pgfplots}
\usepackage{authblk}
\usepackage{subcaption}
\usepackage{adjustbox}
\usepackage{mathtools}
\usepackage{graphicx}
\usepackage{url}

\usetikzlibrary{
    patterns,
    chains,
    backgrounds,
    calc,
    shadings,
    shapes.arrows,
    arrows,
    shapes.symbols,
    shadows,
    positioning,
    decorations.markings,
    backgrounds,
    arrows.meta,
    external
}
\usepackage{array}

\DeclareUnicodeCharacter{2212}{−}
\usepgfplotslibrary{groupplots,dateplot}
\usetikzlibrary{patterns,shapes.arrows}
\pgfplotsset{compat=newest}

\newif\iffinal

\iffinal
  \newcommand{\ian}[1]{}
  \newcommand{\ryan}[1]{}
  \newcommand{\santanu}[1]{}
  \newcommand{\jake}[1]{}
  \newcommand{\maxx}[1]{}
  \newcommand{\zliu}[1]{}
\else
  \newcommand{\ian}[1]{{\textcolor{red}{ Ian: #1 }}}
  \newcommand{\ryan}[1]{{\textcolor{magenta}{ Ryan: #1 }}}
  \newcommand{\santanu}[1]{{\textcolor{purple}{ Santanu: #1 }}}
  \newcommand{\jake}[1]{{\textcolor{blue}{ Jakob: #1 }}}
  \newcommand{\maxx}[1]{{\textcolor{green}{ Max: #1 }}}
  \newcommand{\zliu}[1]{{\textcolor{green}{ Zhengchun: #1 }}}
\fi

\def\BibTeX{{\rm B\kern-.05em{\sc i\kern-.025em b}\kern-.08em
    T\kern-.1667em\lower.7ex\hbox{E}\kern-.125emX}}
\begin{document}
\bstctlcite{IEEEexample:BSTcontrol}

\title{Real-Time Streaming and Event-driven Control of Scientific Experiments}

\author[1]{Jakob R. Elias}
\author[2]{Ryan Chard}
\author[3]{Maksim Levental}
\author[2]{Zhengchun Liu}
\author[2,3]{Ian Foster}
\author[1,2,4]{Santanu Chaudhuri}

\affil[1]{Applied Materials Division, Argonne National Laboratory}
\affil[2]{Data Science and Learning Division, Argonne National Laboratory}
\affil[3]{Department of Computer Science, University of Chicago}
\affil[4]{University of Illinois, Chicago}

\maketitle

\begin{abstract}

Advancements in scientific instrument sensors and connected devices provide unprecedented insight into 
ongoing experiments and present new opportunities for 
control, optimization, and steering. 
However, the diversity of sensors and heterogeneity of their data result in 
make it challenging to fully realize these new opportunities. 
Organizing and synthesizing diverse data streams 
in near-real-time requires both rich automation and
Machine Learning (ML). 
To efficiently utilize ML during an experiment, the entire ML 
lifecycle must be addressed, including refining experiment 
configurations, retraining 
models, and applying decisions---tasks that require an equally diverse array of computational resources spanning centralized HPC to the accelerators at the edge.
Here we present the Manufacturing Data and Machine Learning platform (MDML).
The MDML is designed to
standardize the research and operational environment for advanced data analytics and ML-enabled automated process optimization by providing the cyberinfrastructure to
integrate sensor data streams and AI in cyber-physical systems for in-situ analysis. To achieve this, the MDML provides a fabric to receive and aggregate IoT data and simultaneously
orchestrate remote computation across the computing 
continuum.
In this paper we describe the MDML and show how it is used in advanced manufacturing 
to act on IoT data and orchestrate distributed ML
to guide experiments.

\end{abstract}

\section{Introduction}

The explosion of data generated by next-generation scientific experiments 
necessitates the use of Machine Learning (ML) and Artificial 
Intelligence (AI)~\cite{stevens2020ai} to derive insights 
in a timely manner. This is particularly evident 
in advanced research laboratories that employ a diverse array of
connected devices to monitor experiments, such as temperature and 
flow sensors, optical imaging, spectroscopic instruments, and detectors~\cite{shumate2018iot},
which promise to transform on-the-fly experimentation by enabling the 
concurrent measurement of many experimental characteristics in real-time. 
Integrating heterogeneous scientific sensor data presents exciting new opportunities 
to attain previously unattainable insight into the state of an ongoing 
experiment to detect rare-events and steer and optimize experiments 
as data are collected. 
The gain in efficiency and possibility of scientific discovery depends on the ability to integrate data streams into actionable knowledge or train increasingly smarter unsupervised machine learning models to connect the latent space and automate design of experiments. Autonomous laboratories will invariably depend on trusted data sets and tools. Building of open source architectures for integration of disparate and asynchronous data sets from laboratories to manufacturing platforms will require access to data sets for realistic use cases and streaming data analysis tools in the open source communities.

The adoption of Internet of Things (IoT) devices, sensor networks, industrial IoT tools, and event-driven data management has been particularly widespread in advanced manufacturing~\cite{mourtzis2016industrial, en14123620} and is instrumental to Industry 4.0~\cite{xu2018industry}. Examples of industrial IoT include devices to monitor instruments and shop floors~\cite{mourtzis2016cloud}, 
predict necessary maintenance~\cite{kanawaday2017machine}, and optimize 
processes~\cite{mourtzis2016cloud}.
However, the many different sensors and devices produce 
data streams with widely disparate data types, rates, volumes, and velocities, requiring new techniques and technologies 
to dynamically aggregate and harmonize data
in order to correlate rare-events and make actionable decisions.

Machine Learning (ML) can effectively process large data streams~\cite{marjani2017big}, steer experiments~\cite{pan2021flame}, and 
integrate heterogeneous data from multiple devices~\cite{xue2019deepfusion}.
However, training, deploying, and using ML models 
can require the use of specialized methods, software, environments, and hardware~\cite{dlhub}.
Further, the deluge of data generated by scientific devices and 
the addition of new sensors can quickly exceed the processing 
capabilities of computing resources collocated with experimental apparatus. 
Indeed, achieving near-real-time analysis for online feedback may rely on the use of devices throughout the entire computing continuum, from rapid analysis at edge devices and GPU accelerated laboratory servers, to remote high performance computing (HPC) systems and specialized ML accelerators~\cite{liu2021bridge}.

\begin{figure*}[h]
 \centering
 \includegraphics[width=.95\textwidth,trim=0in 0in 0in 0in,clip]{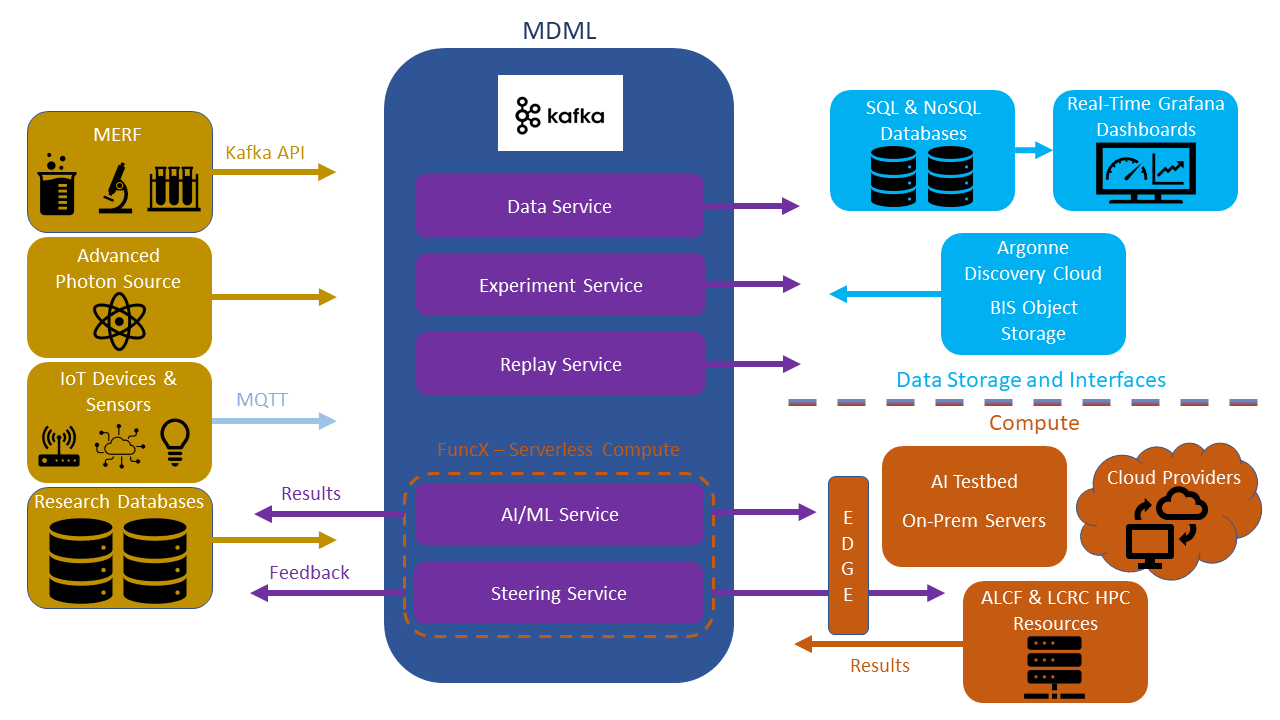}
 \caption{Manufacturing Data and Machine Learning platform provides services for scientists to extend their capabilities in data streaming, online analysis, experiment steering, and real-time monitoring and visualization.}
 \label{fig:mdml-overview}
\end{figure*}

In this paper we present the Manufacturing Data and Machine Learning (MDML) platform to 
address these challenges. The MDML, depicted in~\figurename{~\ref{fig:mdml-overview}}, provides foundational cyberinfrastructure to standardize the research and operational environment to act on scientific IoT and instrument data streams and integrate ML in the loop to steer experiments. The MDML facilitates advanced data analytics and enables
automated, ML-driven optimization of experiments. The MDML has been designed 
to meet the needs of in-situ measurements for accelerating scalable materials manufacturing while providing capabilities that can be easily adapted to any scientific domain.
The MDML enables users to construct rich, data-oriented analysis pipelines that span disparate computational environments, from laptops and local servers to supercomputers and clouds, and provides visualization and monitoring tools to create dynamic interfaces to experimental facilities and their processing pipelines.

The MDML platform (Fig. 1) is collection of multiple services serving multiple streaming data sources with ability to deliver real-time analysis and steering of materials processing and manufacturing systems. The user can use simple Python API and assemble their own publish/subscribe architecture for running MDML use case. On the left of Fig. 1, the data sources can be streamed from various device or traditional databases with very little change to their current data model and protocols. As a Pub/Sub model, functions on the left is only responsible for publishing. On the right, MDML can send these data streams to consumers that can scale depending on the need for real time feedback, size of data, and complexity of the compute functions. The ability to scale is provided by serverless compute capabilities in MDML. MDML also has data service, experiment design and execution service to integrate control system hardware, replay services using data-fusion functions, and capabilities to train and deploy AI/ML models as service to serve autonomous or human-in-the-loop functions.  The Fig. 1 also shows how assets deployed in the ANL's materials engineering research facility (MERF) can connect ability to get real time data sets from Advanced Photon Source (APS) and send the heavy calculations to Argonne Leadership Computing Facility (ALCF). However, MDML can run fully at the local level in just connecting custom sensors and IoT modules using local Edge and GPU-based resources. The MDML platform being light can span and connect edge-to-exascale computing continuum as our vision for the future. MDML is fully flexible in making the choice between publishers, data ingestion rates, and real-time response performance have multiple dependence on hardware, complexity of the problem, and quality of data sources. 

We evaluate the MDML's capabilities using three real-world experimental use cases---combustion synthesis, electrospinning, and a continuous flow reactor. Each presents
a data-intensive experiment employing multiple sensors.
We describe these experiments and their requirements and explain
how the MDML has been applied to facilitate on-demand
analysis for dynamic optimization and steering.
The experiments leverage a range of distributed computing devices, including edge accelerators co-located with the instruments, GPU-accelerated laboratory servers, and HPC facilities. 

This paper extends our preliminary work~\cite{elias2019mdml} and presents the following major contributions. 
\begin{itemize}
    \item Describes the design and implementation of the MDML.
    \item Highlights the unique requirements of next-generation laboratories and advanced manufacturing.
    \item Presents an evaluation of the MDML's capabilities to serve ML inference across the computing continuum.
    \item Applied MDML to three real-world manufacturing use cases.
    \item Discusses how the MDML is used to facilitate near-real-time monitoring and feedback in three real-world experiments.
\end{itemize}

The remainder of this paper is organized as follows. \S\ref{sec:manufacturing} discusses the requirements of advanced manufacturing. \S\ref{sec:mdml} describes the MDML architecture and implementation. \S\ref{sec:eval} presents an evaluation of the MDML and \S\ref{sec:control} describes its application to various use cases. Finally, \S\ref{sec:related} presents related works and concluding remarks are given in \S\ref{sec:conc}.

\section{Advanced Manufacturing Requirements}
\label{sec:manufacturing}

The MDML is designed to meet the needs of modern research laboratories by
enabling scientists to dynamically
integrate heterogeneous data sources 
to guide and 
optimize experiments. 
Argonne National Laboratory's Materials Engineering Research Facility (MERF) is one such example, housing a variety of high-throughput, data-intensive instruments for 
scale-up manufacturing. 
These instruments include myriad sensors and cameras to monitor the 
instrument, samples, and phenomena during an experimental campaign.
Here we briefly describe three MERF experiments and highlight their data and compute requirements.

\emph{Flame Spray Pyrolysis} (FSP) is a combustion synthesis reaction to create nanomaterials from
high heat of combustion liquids dissolved in a solvent~\cite{SOKOLOWSKI1977219}. The FSP instrument employs a Planar Laser Induced Fluorescence (PLIF) diagnostic system that uses a tunable laser light sheet to characterize flame chemistries every 50ms, spectroscopy to determine the contents of the exhaust every few minutes, and measures of the particle size distribution for the resulting nanomaterials. 
Scientists using the FSP instrument rely on a laboratory server with GPU accelerators to process PLIF images and a connected computer to view spectroscopy results as they are generated.

\emph{Electrospinning} is a manufacturing method that produces nanometer-scale fibers 
for use in the production of materials ranging from batteries to filters. 
Using electrical fields, electrospinning instruments draw charged threads of
polymer into unwoven strands of fiber.
A crucial step in this process is to detect defects, such as large droplets of material rather than strands of fiber and modify the instrument's configuration to optimize the manufacturing process. To detect these defects MERF scientists leverage a high throughput, GPU-enhanced detection code.

\emph{Continuous flow chemistry} enables the synthesis of materials using a flowing stream of reactants to perpetually produce a stream of products as an experiment runs. Continuous flow reactors enable scientists to synthesize materials from discovery through process development and production in a cost effective manner. Continuous flow reactors differ from traditional, batch synthesis techniques, whereby products are typically created by combining materials in a silo to produce batches of outputs. Such techniques are often unable to scale the manufacture of novel materials cost-effectively.

\subsection{Requirements}

We use the experiments described above to derive the requirements for the MDML and summarize them below.

\emph{Data management:} Each of the three experiments employ a range of sensors that produce data at different rates and volumes and require high throughput, low latency data management and streaming solutions. 
In addition, while some sensors generate discrete datasets, such as the PLIF images created during FSP experiments, others generate continuous streams of data that provide opportunities for continuous analysis, such as flow rate monitors in the continuous flow reactor. Therefore, a common data management solution must accommodate both discrete data events and data streams.
Further, the experiments employ distinct data storage solutions, including databases, on-premise storage, and cloud-based object storage. Therefore the data management requirements include capabilities to connect to external storage devices to persist data.


\emph{Flexible computing resources:} Scientists require capabilities to compute wherever most suitable---where a suitable computer
is available, software is installed, or data are located, for example. This concept is far from new~\cite{parkhill1966challenge}, and motivated initiatives such as grid~\cite{foster2001anatomy} and peer-to-peer computing~\cite{milojicic2002peer}. 
This concept is particularly important in laboratory settings where an array of computing devices are readily available to a scientist, from edge accelerators and personal laptops, and laboratory servers, through to clusters and HPC resources~\cite{vescovi2022linking}. Further, high speed, reliable networks interconnect the instruments with the computing continuum, enabling the exploitation of any available computing resource. Therefore, the MDML should enable users to perform analyses across a wide range of heterogeneous computing resources. 
Finally, it is important for the MDML to facilitate the use of data to be used when creating visualizations and portals. These portals are critical to providing understanding into an ongoing experiment and necessitate that the MDML be capable of making data feeds available to users for visualization purposes.

\emph{Connectivity and interoperability:} Each of the use cases highlight the need to integrate with established software and hardware stacks, from proprietary instrument software such as LabView through to legacy MatLab interfaces. To accommodate such a requirement the MDML should provide abstractions to integrate existing software and sensors.
Further, the above use cases are well established with a rich collection of tools and transformation pipelines that would be time consuming and costly to modify or replace. Thus, the MDML must accommodate the application of custom analysis tools and transformations to data.

\emph{Optimization and steering:} The multitude of sensors associated with a given instrument creates a large, multidimensional optimization problem. When combined with the different data types, rates, and volumes created by different sensors, detecting anomalies and automatically optimizing an experimental process is untenable for human users and necessitates the use of ML models.
Thus, it is critical for the MDML to enable scientists to deploy and integrate ML models into the experimental process. The MDML must accomodate
the deployment of models and enable their application to data streams.
Further, during the creation of production-quality ML models, there is often a need to compare several models with varying hyper-parameter tunings to find an optimal solution. The MDML should assist these efforts by facilitating the replay of data events that occurred during a real experiment, such that
multiple models can be tested concurrently using hand-picked experiments to evaluate their ability over a range of experimental conditions.

\section{The Manufacturing Data and Machine Learning Platform}
\label{sec:mdml}

\begin{figure}[h]
 \centering
 \includegraphics[width=\columnwidth,clip]{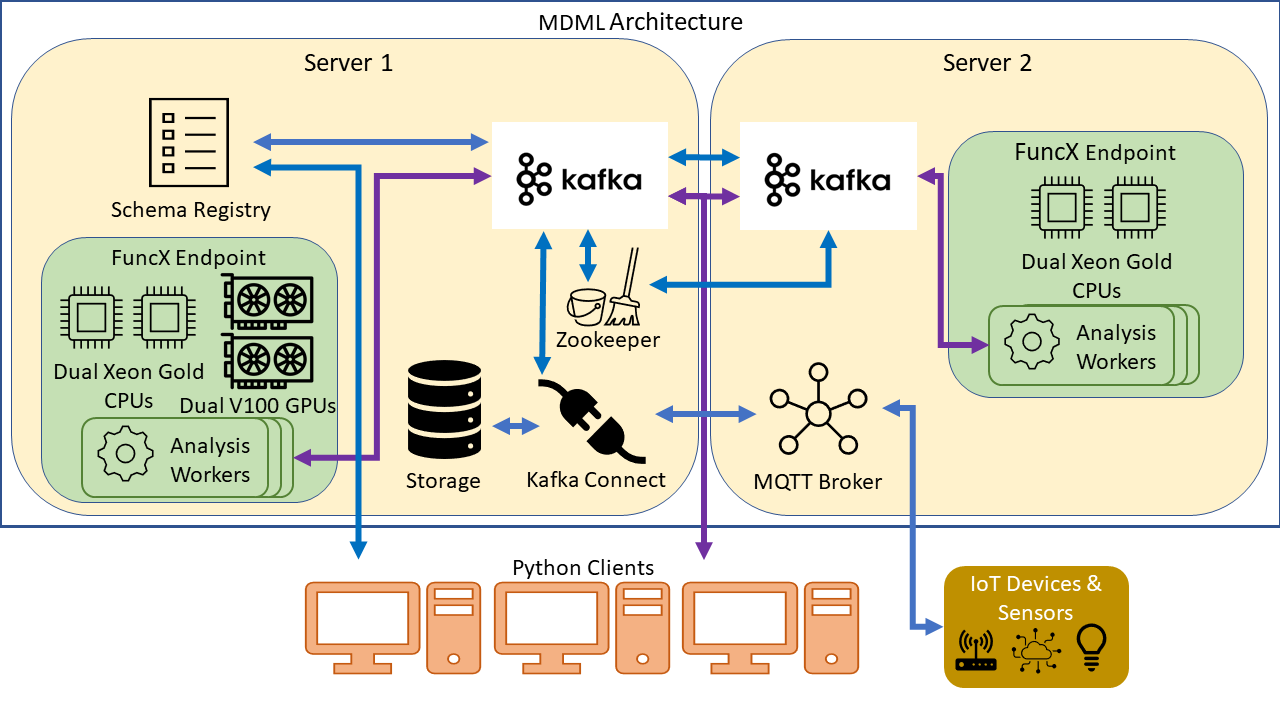}
 \caption{MDML architecture with Kafka and funcX}
 \label{fig:mdml_arch}
\end{figure}

The MDML, depicted in~\figurename~\ref{fig:mdml_arch}, enables 
the optimization and steering of scientific experiments by 
connecting IoT assets and their data streams to analysis capabilities throughout the computing continuum. 
To achieve this, the MDML provides capabilities to publish sensor data, manage and organize data streams, persist data through multiple storage connectors, orchestrate remote consumers to perform analysis, return results to the instrument, and collect and replay experiment data.
Here we describe the architecture and implementation of the MDML.

\subsection{Architecture}

The MDML is designed around two key components: The hosted \emph{MDML Service} that receives and manages instrument data streams, and remote \emph{MDML Agents} that produce and consume data. The service hosts a set of \emph{Topics} for agents to exchange data with the service and other agents.

\subsubsection{MDML Service}

The MDML service is the central broker for the MDML, facilitating the connection of users, instruments, storage, and compute. Data are securely published to the service where they are marshalled to their appropriate destination, whether that be storage through an 
appropriate connector, or for analysis by a consuming agent.
The service also maintains a registry of devices, experiments, and clients, and exposes interfaces for new devices to integrate into the system. When a device is registered with the system a new topic is provisioned to receive data. These topics can be aggregated by the service and consumed by other agents.

The service integrates a multitude of data storage technologies to organize and persist experimental data. These stores include time series databases to capture data streams and events, object storage for storing large datasets, and relational databases to maintain relationships across data streams. 
These storage capabilities enable MDML users to aggregate and construct \emph{experiments}, where collections of data events are stored in a fashion that allows them to be replayed on-demand, whereby the events of the experiment are redelivered to the topics, emulating the experiment and establishing a digital twin that can then be used to test and evaluate different analysis approaches. 

\subsubsection{MDML Agents}

MDML agents are deployed at a remote resource and are responsible for communicating with the MDML service for data publication and retrieval. The agent is designed to be installed alongside an instrument or sensor and act as an interface to the MDML. As such, the agent software is designed to be light-weight and easily installable, and 
introduce minimal overhead. 
Once deployed, an agent must register with the MDML service before it can begin publishing or receiving data. The agent can be customized with instrument-specific data schemas to control how data are acted upon and stored once received by the service. These schemas may also contain metadata about the data being streamed to aid new users of the data. 

The agent provides several capabilities for users to work with their data streams:
\begin{itemize}
    \item Stream and store data: By streaming data via the MDML client, data is not only saved in the MDML service but also any peripheral data storage solutions in place: time series databases, persistent object storage, and more.
    \item Metadata tagging: User-defined schemas are associated with data streams to provide insight about the type of data. Schemas also enable consumers to automatically store data in services external to the MDML service like SQL databases.
    \item Consume streams for analysis: Users are given the location of a topic that can be passed to an agent to consume data as they are published to the service.
    \item Aggregate streams: Users can specify multiple data streams to consume from - effectively fusing data from multiple sources/devices.
\end{itemize}

In addition to publishing data, the agent software provides similar capabilities to retrieve data from the service. This model is used when deploying analysis tools to act on data as they are produced (e.g., ML inference, image processing, and data storage). These results can then be produced back to an experimental setup via the MDML service to facilitate online steering and optimization.

\subsection{Implementation}

The MDML is built upon Apache Kafka~\cite{kreps2011kafka}, funcX~\cite{funcx}, Globus~\cite{chard2016globus}, and a host of data storage solutions to provide a reliable and secure cyberinfrastructure to stream, manage, and analyze data for the development of optimization and control systems.
Here we describe how these components are employed and their roles
in the MDML.

\subsubsection{MDML Service}

The MDML service is deployed using Docker Compose to create a cluster of 
services that automatically handle the partition and replication of data 
streams. This gives the MDML greater availability and fault tolerance. The primary deployment of the service is hosted on premise at Argonne National Laboratory in the Materials Engineering Research Facility but it is also available as a cloud-hosted service.

We leverage Apache Kafka to facilitate high-performance, reliable data streaming between remote agents and the MDML service. Kafka is an open-source, event-based streaming platform that enables the sending of arbitrary sensor and instrument data, from small sensor streams, through to large images.

Each device engaging the MDML is provisioned a unique, namespaced Kafka topic on which to publish data. These topics serve as queues of data events that have occurred during experiments. Data consumers can then access a topic to consume data that are published.
Consumers can be configured to read messages from the beginning of a topic, the last message received, or from an arbitrary point in the topic. A consumer can read from a topic indefinitely, be configured to retrieve a certain number of messages, or read until a timeout period has been reached with no incoming messages.

Multiple consumers can be started to consume data from the same topic if needed. For example, when data generation is faster than the analysis task, multiple instances of the analysis can be used to simultaneously pull from the same data topic. This allows the analysis tasks to scale up with the rate of data generation. Furthermore, by specifying the same group ID for each consumer, Kafka automatically handles sharing data between members of the group so that messages are received only once.

\begin{figure}[h]
 \centering
 \includegraphics[width=\columnwidth,clip]{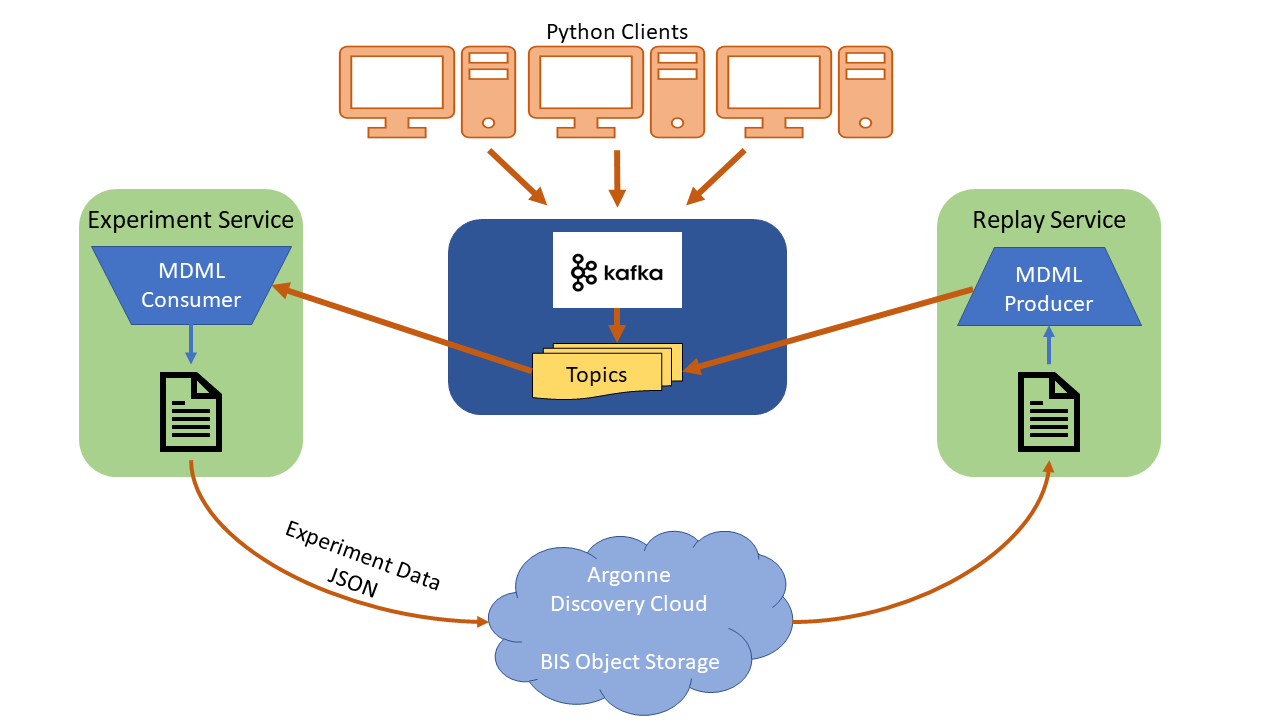}
 \caption{During an MDML experiment, topics defined by the experiment are fed to a consumer that will record the messages streamed during an experiment. Upon replaying the experiment, a file that was created and saved containing the experiments data is fed to an MDML producer. This producer then publishes the same data messages to their original topics. 
}
 \label{fig:mdml_replay}
\end{figure}

\subsubsection{MDML Agents}
MDML agents are responsible for transmitting, retrieving, and acting upon data streams. We have developed a Python toolkit to simplify both the creation and deployment of MDML agents. Our toolkit 
abstracts communicating with the MDML service by wrapping the Confluent Kafka Python API. 
Our toolkit removes superfluous details related to Kafka that should not be a concern for domain scientists. 
The toolkit also includes functionality for streaming large messages in chunks as there is a maximum message size in Kafka, a ``coat check" system that uses Kafka and S3 buckets for transferring large files, automated JSON schema creation, and controls for the experiment and replay services.

In addition to producer and consumer agents, users can deploy Kafka Connectors to act on data streams. 
Connectors are named as such as they connect Kafka topics to other storage solutions (e.g., SQL databases, AWS S3, MongoDB, and Redis) and message passing protocols (such as MQTT). 
Many connectors have been developed by the Kafka community to perform common tasks and are widely available through the community hub for Kafka Connectors. These connectors can be easily employed to act on MDML data streams and allow users to configure custom storage solutions. 

To deploy and manage Kafka Connectors the MDML leverages the Kafka Connect service.
This means the user simply defines a connector's configuration and the service will deploy and host the connector. A connector's configuration describes which topic and data storage solution or protocol to connect as well as any relevant connection details. Kafka Connectors can act as either a sink or a source. Sink connectors push data from a topic to storage solutions while source connectors pull and monitor changes from a storage device to send data out to Kafka topics. Source connectors assist in the adoption of the MDML by allowing easy ingress of data from existing sources or legacy systems. On the other hand, sink connectors allow researchers to adopt the event-driven data practices provided by the MDML while continuing to access their data through familiar data storage systems of their choosing.

\subsection{Data Management}

Data stream events are logged to files corresponding to a given topic. As a result, all data is persisted and is ready to be retrieved by consumer agents on-demand. 
In addition to these log files, the Kafka Connectors mentioned above allow data to be stored in nearly any commonly used data storage solution. 

An experiment is defined as a list of topics that will have data produced to during a specific scientific experiment. The user has control over the start and end of the experiment. Between the start and stop calls, all messages streamed to the given topics are consumed, written to a file, and subsequently uploaded to the Argonne Discovery Cloud ~\cite{argonne-discovery-cloud} for sharing with others or to facilitate replaying an experiment.
Replay capabilities are achieved by re-streaming 
all of the messages that were received during an experiment. Using the temporal information attained during 
the original collection, the messages can be reproduced 
with the same order and timing that occurred in the experiment. This enables users to develop codes or test new analyses by running digital twin replays of real-world experiments. Replays use the same topics as the original experiment to seamlessly transition between live and emulated data.
Furthermore, due to the use of the Argonne Discovery Cloud to host experiment files, an experiment can be replayed on any deployment of the MDML, regardless of the MDML deployment that originally captured the experiment.


\subsection{Visualization}

MDML provides dashboards as a means to visualize and monitor an experiment's data.
MDML employs the industry standard Grafana dashboard as a customizable platform to create dynamic and interactive visualization experiences. Users can configure their own dashboards to visualize sensor measurements, the state of analysis tasks, and results as data are processed. This not only allows researchers to dissect their data in real-time but also acts as a monitoring tool for critical experimental data. By taking advantage of Kafka Connectors, a PostgreSQL database is used to store data messages streamed to the MDML under specific topics. These topics are given their own data tables in PostgreSQL. Grafana, in turn, periodically queries the PostgreSQL database to update the dashboards visualizations to near real-time.   





\section{Evaluation}
\label{sec:eval}


We evaluate the MDML in terms of throughput and scalability and review its ability to leverage distributed resources to perform tasks. Finally, we review the MDML's application to the three real-world use cases described in \S{\ref{sec:manufacturing}}.

\subsection{Experimental Setup}

The MDML is hosted in the Materials Engineering Research Facility (MERF) at Argonne National Laboratory. 
The MERF has a 10 gigabit fiber network with connections in all of the lab spaces. Two HPE DL380 servers located in the MERF are used to host the MDML service. These servers have dual Intel Xeon Gold CPUs. One of the servers hosts two NVIDIA Tesla V100 GPUs. 
In addition to the MDML service hardware, the MERF also contains various edge nodes. There are four NVIDIA Jetson AGX Xaviers, an NVIDIA Jetson TX2, and an UP-Xtreme edge computing kit. These edge nodes are used for various purposes such as simulating and testing client jobs or offloading analysis tasks.
A second deployment of the MDML is hosted using an Amazon EC2 t2.xlarge instance in the US-East-1 region. 

\begin{figure}[h]
 \centering
 \includegraphics[width=\columnwidth,trim=0in 0in 0in 0in,clip]{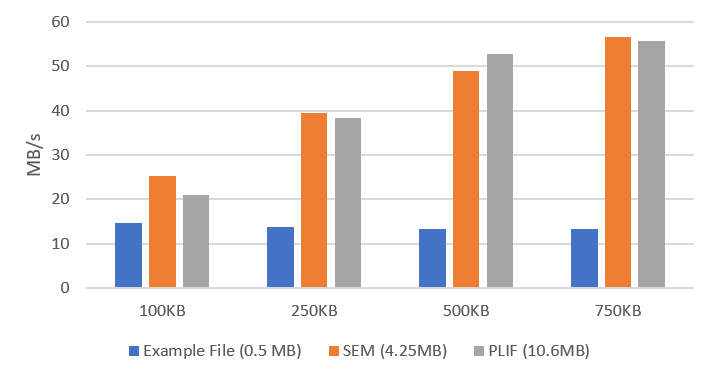}
 \caption{Data rates by chunking large messages.}
 \label{fig:rates_by_chunk}
\end{figure}

We first evaluate the MDML's ability to stream scientific data of different sizes.
As mentioned earlier, Kafka imposes a maximum size for messages. The MDML configures this maximum size to be 1MB for each message. However, there are many use cases in which messages or files larger than 1MB will be streamed. To accomplish this, the MDML python client implements a method of chunking larger messages into parts which are then automatically reassembled upon a client consuming the messages. \figurename~\ref{fig:rates_by_chunk} shows the MDML's data streaming rate in MB/s when using various chunk sizes. Unsurprisingly, we see an example file of 0.5MB does not take advantage of chunking. However, when streaming SEM and PLIF images of 4.25MB and 10MB respectively, the maximum data rate increases as the chunk size increases. This information allowed us to configure the MDML client to use the largest chunk size available to maximize data transfer rates when chunking is required.

\begin{table}[H]
    \centering

    \begin{center}
        \begin{tabular}{| l | l | l | l |}
        \hline
         & 1 Worker & 2 Workers & 4 Workers \\ \hline
        Analysis Rates (MB/s) & 11.67 & 23.50 & 38.72 \\
        \hline
        \end{tabular}
    \end{center}
    \caption{Electrospinning image classification data rates with varying numbers of FuncX workers}
    \label{tab:espin}
\end{table}


Next, we examine the platform's ability to scale analysis tasks using multiple consumer agents.
\tablename~\ref{tab:espin} shows the result of scaling an image classification task to detect defective fibers from an electrospinning experiment. We leverage the FuncX function-as-a-service platform to start varying numbers of consumer agents. The figure shows the total amount of image data being classified in terms of MB/s. FuncX was used to create 1, 2, and 4 concurrent consumer agents to perform the analysis task. Each consumer belongs to the same group ID so that each image is processed only once.
This shares the task of data analysis across a configurable numbers of workers. As expected, we see that the analysis rates scale linearly with the addition of new analysis workers. While there is no bound to the number of FuncX workers that could be used, we are limited by hardware. The image classification models were loaded onto Nvidia Tesla V100 GPUs. Since each GPU can only load two instances of the model and the FuncX endpoint's server contains only two of these GPUs, we are limited to 4 FuncX workers for this test. This test demonstrates that regardless of the data generation rate, the underlying mechanism for running on-demand analyses with MDML, namely FuncX, can scale such that analysis tasks keep pace with data generation. 



\section{Experimental Control}
\label{sec:control}

We have applied the MDML to the three real-world scientific use cases described in \S\ref{sec:manufacturing}. These
case studies are representative of data-intensive and compute-intensive manufacturing
approaches that rely on diverse data streams, machine learning for optimization and
control, and leverage resources throughout the computing continuum. 

\begin{figure*}[h]
 \centering
 \includegraphics[width=\textwidth,trim=0in 0in 0in 0in,clip]{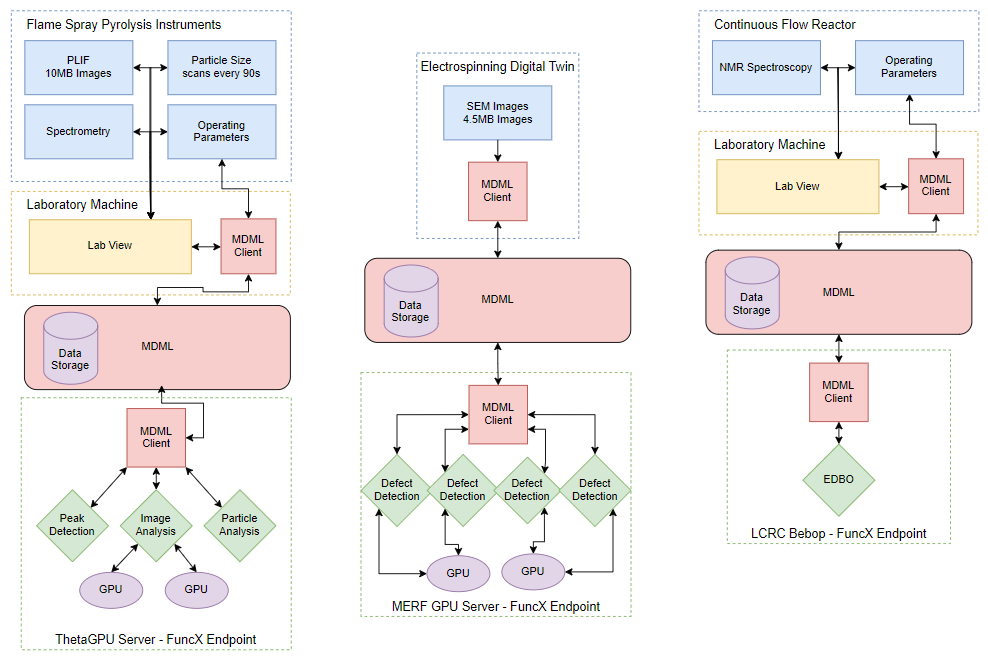}
 \caption{MDML use cases diagram.}
 \label{fig:mdml}
\end{figure*}

\subsection{Flame Spray Pyrolysis}

We used the MDML to orchestrate the analysis pipeline for FSP and perform near-real time quality control of the flame's stability. The MDML provides scientists tools to both monitor the state of the experiment and steer the evolution of the flame. Using local resources for rapid quality control, HPC resources for large-scale analysis, and ML models to integrate diverse data types, The MDML assists in guiding flame stability. Finally, experiments are visualized in near-real-time via an interactive interface using the Grafana platform.

\begin{figure}[H]
    \centering
   \includegraphics[width=.9\columnwidth]{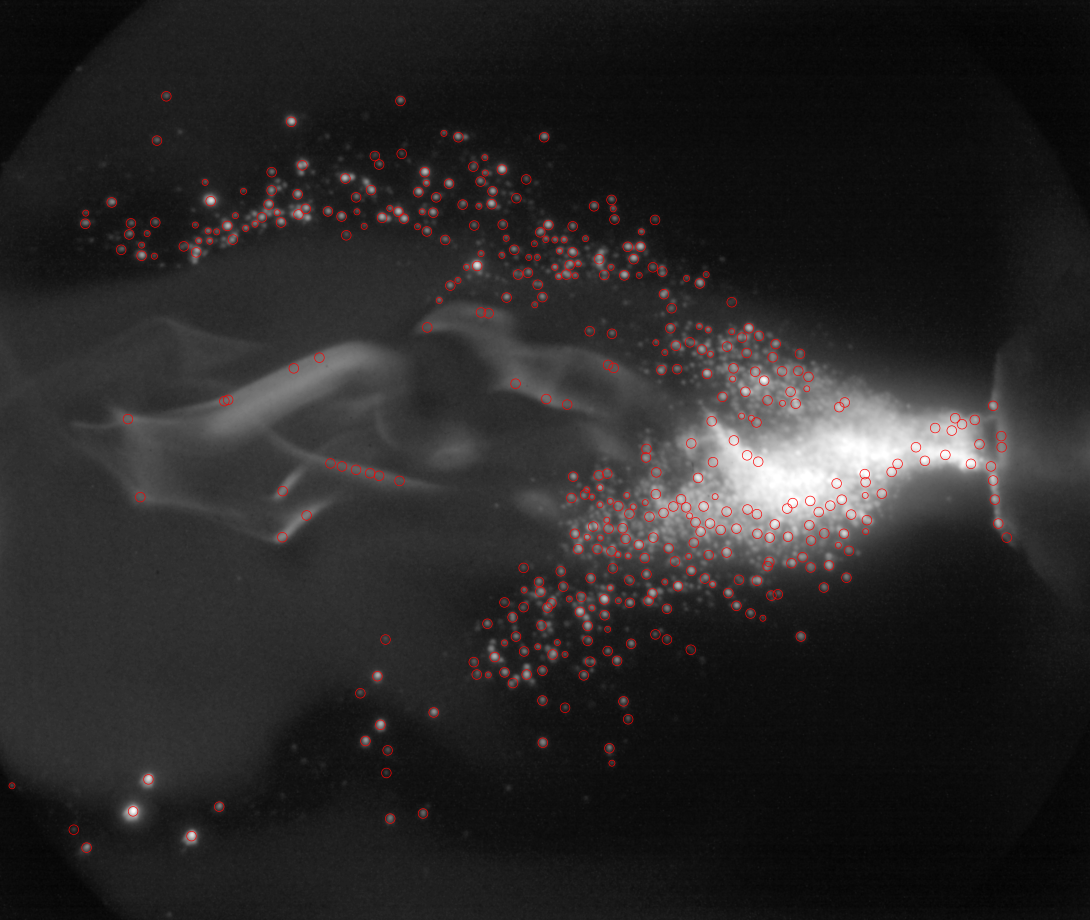} 
    \caption{PLIF instrument sample. Diffuse regions are indeed induced fluorescence, while brighter regions are the result of Mie scattering by poorly atomized fuel droplets. Blobs/droplets detected in the image are identified with red circles}
    \label{fig:plif2}
\end{figure}

To achieve online monitoring of the FSP experiment we extend our prior work in~\cite{merf-fsp-paper} and deploy the analysis tool as an MDML consumer agent. This tool analyzes PLIF imagery collected at 10Hz to infer droplet volume distribution (see figure~\eqref{fig:plif2}).
The PLIF images are passed to an analysis tool which locates droplets and infers their radii.
Blob detection is the identification and characterization of regions in an image that are distinct from adjacent regions and for which certain properties are either constant or approximately constant.
For this work we implemented a novel blob detection tool for GPUs using PyTorch~\cite{NEURIPS2019_9015}; code and documentation are openly available~\cite{merf-fsp-github}. Information regarding the characterization of blobs are then fed back to the instrument where configurations are changed to minimize both the quantity and size of uncombusted fuel droplets.

\input{figures/gpu_vs_cpu}

The performance improvement of the GPU implementation of the blob detector is readily apparent.
The blob detector has approximately constant run-time in \emph{n\_bin} (i.e. the number of blob radii detecting) because of the parallelism of GPU compute (see \figurename~\ref{fig:gpu_vs_cpu}) while the CPU implementation scales linearly.
Note that the discontinuous increase in runtime at $\mathtt{n\_bin} = 32$ is due to CUDA's internal optimizer's choice of convolution strategy. This allows the MDML to orchestrate the analysis of PLIF images faster than they are generated.

\subsection{Electrospinning}

The MDML has also been applied to the MERF's scale-up electrospinning
experiments which are guided by analysis and process optimization performed at the Advanced Photon Source (APS). In electrospinning, a polymer solution is charged and sprayed on a support to make non-woven deposition of nanoscale fibers. The diameter and distribution of fibers are dependent on velocity of liquid jets, their drying on the air, and charged states when exiting the spray needle or nozzle. A large roll-to-roll version of electrospinning has 56-nozzles and the APS version has a single one to tune the parameters for the process. The APS version has an automated setup that can spin fibers in the path of the x-ray beam.

\begin{figure*}[h]
    \centering
   \includegraphics[width=.9\textwidth]{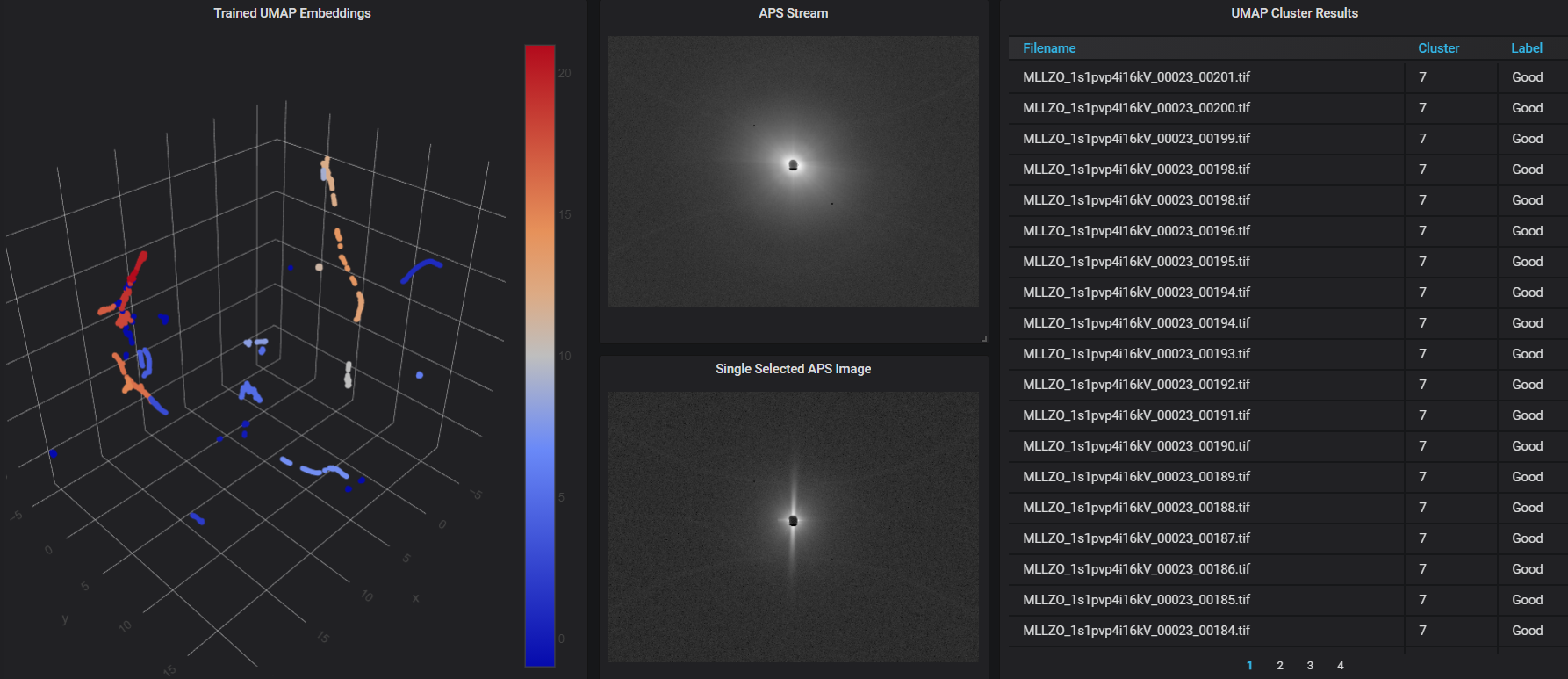} 
    \caption{UMAP embedding and clustering of SAX images.}
    \label{fig:umap}
\end{figure*}

Using the MDML as a bridge, feedback from the small-scale experiment can be applied to tune and guide the large-scale production of fibers.
We used Uniform Manifold Approximation and Projection~\cite{mcinnes2018umap} (UMAP) and HDBSCAN~\cite{hdscan} in order to cluster training SAX images of the electrospinning products according to their quality/fitness. 
We then performed cluster assignments in real time to determine whether the electrospinning process trajectory was tending towards high quality fibers. 
All of the data streams and inferences were orchestrated by MDML (see~\figurename{\ref{fig:umap}}). 
The size of the data generated during this experiment were 1.2MB TIF images. As the images are output from the beamline, they are streamed to the MDML at 24 MB/s. While each beam line at the APS has varying data sizes and rates, the underlying mechanism that allows streaming via MDML is available at every beam line. Thus, while this is only one experiment at one beam line, the larger workflow of streaming to MDML and consuming the data within a FuncX task for analysis can be generalized across the APS' beam lines.

\subsection{Continuous Flow Reactor and design of experiments}
The continuous flow reactor (CFR) at Argonne presents an exciting opportunity to realize a self-driving laboratory, where experiments are performed autonomously. To accomplish this, the CFR leverages the MDML to integrate an Experimental Design using Bayesian Optimization (EBDO) tool to optimize the reaction yield \cite{edbo-paper}. EDBO takes in discrete lists of all experiment parameters and selects combinations of parameters that have a chance to maximize (optimize) the amount of yield from the reaction. The CFR controls settings, data output, and analysis code all communicate with the MDML service to exchange experiment results and newly suggested experiment settings. In this way, the operation of the experiment becomes fully autonomous as the EDBO code prescribes new experiments for the machine to run. The in-line NMR data right now uses commercial databases and line-fitting to understand location of peaks. Rapid prediction of NMR shifts with uncertainty quantification is now possible with accuracy close to first-principles methods \cite{jonas2019rapid}. As a next step, predictions of NMR peaks and their changes can be fully automated by streaming the raw NMR data during synthesis to MDML. The use of quantum chemical calculations for predicting reaction products and relative stability also currently depend on solvation models which are not ready for complex multi-solvent mixtures for synthesis and the recalculation of kinetics under temperature-pressure-flow conditions available in the CFR. In heavily instrument CFR, multi-modal measurements, and streaming of asynchronous data through MDML are less useful. With integration of physics-based AI, automated synthesis design tools, and growing list of hyperparameter training capabilities, the events detected by MDML  can be analyzed and related to chemical transformations. Therefore, MDML promises to be an important event-driven scientific AI platform for merging multi-objective optimization in chemical synthesis and discovery of new pathways to novel end-products.  
\begin{figure}[H]
 \centering
 \includegraphics[width=\columnwidth,trim=0in 0in 0in 0in,clip]{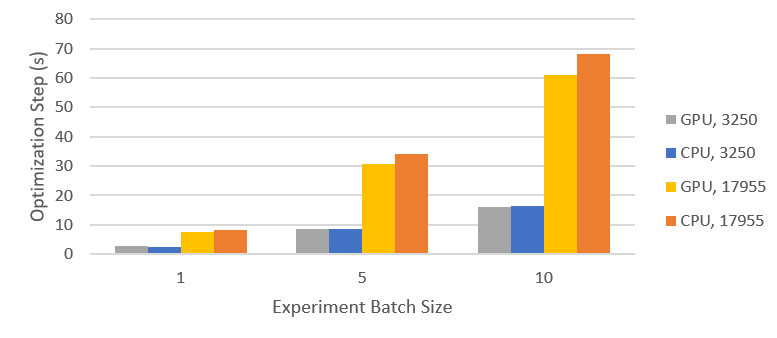}
 \caption{Time to complete the Experimental Design via Bayesian Optimization (EDBO) optimization step on CPU vs. GPU, with variable batch size and design parameter space}
 \label{fig:edbo}
\end{figure}

\figurename~\ref{fig:edbo} shows the experimental design via Bayesian optimization (EDBO)~\cite{edbo-paper} analysis run for sets of 100 experiment suggestions with a few key differences between each set: the total number of possible experiment settings to search over, the batch size of suggested experiments, and the hardware used to run the analysis - GPU vs CPU. The time being measured is the amount of time the analysis takes to complete the optimization step. This is the amount of time that the CFR machine would be idle waiting for a new batch of suggested experiments. The results show that 
increasing the number of possible parameter combinations increases the time for the EDBO algorithm to identify new experiment settings. The same is true for the batch size - the number of experiments suggested after each optimization step. However, since the increase in analysis time from an increase in batch size is small, it is better to run with a larger batch size than more iterations with a smaller batch size. 
To ensure these experiments were consistent and representative of the real-world instrument, we performed the tests using the MDML's replay capabilities.

During the writing of this paper, work is underway to integrate Argonne's Parallel Multi-Objective Optimization code (ParMOO) \cite{parmoo} with MDML. ParMOO can act as a replacement for EDBO with the bonus of allowing for continuous parameter spaces which fits better in the CFR experiment setting. However, ParMOO does not take the underlying solvent's chemical properties into account while solving as EDBO does. We plan to explore these implications once integration is complete.


\section{Related Work}
\label{sec:related}

There is an abundance of work surrounding event-driven data management in the context of manufacturing processes~\cite{doi:10.1080/23307706.2014.885288}. One such example uses Kafka, Grafana, and SQL databases to collect, monitor, and act on manufacturing data~\cite{en14123620}. The main difference with their approach is using the ksqlDB database to perform operations on data. The MDML uses FuncX, a serverless compute platform, to create deployable functions that act on data. This allows streaming data in the MDML to be utilized by advanced machine learning methods that are not available in ksqlDB.

This paradigm shift to handling event streaming data requires new machine learning methods that traditionally operated in batch or offline systems. Gomes et al. ~\cite{10.1145/3373464.3373470} explores the challenges and current solutions for building machine learning models in an event-based, streaming architecture. One particular solution studied is a python package called scikit-multiflow. This package builds on the popular machine learning package scikit-learn. The MDML allows users to take advantage of this package and many more through the function serving platform FuncX. In this way, MDML users can still leverage state of the art ML codes.  

With the rise of the Internet of Things (IoT) several new protocols have been created to handle the exchange of information between devices~\cite{Happ2017MeetingIP}. One of the more popular protocols, MQTT, has dozens of broker implementations that provide various features on top of the basic publish/subscribe model that the protocol provides. Established scientific experiments may be locked into using IoT protocols. The MDML makes it possible to integrate with existing MQTT installations via Kafka Connectors. With a properly configured connector, messages published via MQTT would be automatically produced down a corresponding Kafka topic and be available to all of the discuss features of the MDML.

In ~\cite{SAHAL2020138}, the authors highlight numerous technologies for running predictive maintenance on big data problems in railway transportation and wind turbines. Kafka is one of the streaming softwares they included in their review. This reaffirms that the MDML is capable of solving real-time optimization problems while scaling to meet the demands of big data seen in manufacturing and other sectors.






\section{Conclusion}
\label{sec:conc}

The emergence of IoT devices and 5G enabled devices presents new challenges and requirements for real-time data streaming applications and their coupling with machine learning. While especially true for manufacturing to allow predictive maintenance, machine steering, and anomaly detection, the problem is common across almost all scientific domains. The MDML is designed to address this need, providing capabilities to integrate connected streaming devices with analysis capabilities.
The addition of new methods for data fusion, algorithm testing using multiple data streams, dynamic testing of ML models, and the ability to replay experiments and compare and benchmark analysis techniques and ML models presents profound opportunities. Here we presented the MDML and demonstrated the ability for it to effectively address these challenges.
We showed the MDML can efficiently accommodate diverse data streams and dynamically scale analysis resources on-demand, using the FuncX platform. We also showed how the system can incorporate specialized hardware and deploy highly optimized codes for real-time analysis. The MDML provides intuitive and flexible interfaces and dashboards. Finally, we evaluated the MDML's application to three real-world manufacturing experiments and showed it can accommodate the diverse requirements of experimental science applications.

In future work for the MDML, we want to increase the security of the platform. Encrypting client connections to Kafka for message streaming was not enable in favor of better data transfer rates between producers and consumers. This feature should be implemented before future external deployments of the MDML. We would also like to add access control lists to topics so that only approved clients can stream data to specific topics. 
In addition to security, we aim to provide user interfaces to review data published in the MDML. We are also actively exploring tools to provide administrative dashboards for users to inspect and manage connected devices, their topics, and consumers of those topics. Finally, we aim to productionize the MDML's deployment and make it available to the wider scientific community.

\section{Funding Acknowledgment}
\label{sec:fund}

This work was funded by the Manufacturing Science and Engineering Initiative at Argonne National Laboratory and by a Laboratory Directed Research and Development (LDRD 2021-0161) project for Real-time IoT Fusion for ML-Driven Experiment Steering at Argonne National Laboratory.

\bibliographystyle{IEEEtran}
\bibliography{refs}

\begin{thebibliography}{10}
\providecommand{\url}[1]{#1}
\csname url@samestyle\endcsname
\providecommand{\newblock}{\relax}
\providecommand{\bibinfo}[2]{#2}
\providecommand{\BIBentrySTDinterwordspacing}{\spaceskip=0pt\relax}
\providecommand{\BIBentryALTinterwordstretchfactor}{4}
\providecommand{\BIBentryALTinterwordspacing}{\spaceskip=\fontdimen2\font plus
\BIBentryALTinterwordstretchfactor\fontdimen3\font minus
  \fontdimen4\font\relax}
\providecommand{\BIBforeignlanguage}[2]{{%
\expandafter\ifx\csname l@#1\endcsname\relax
\typeout{** WARNING: IEEEtran.bst: No hyphenation pattern has been}%
\typeout{** loaded for the language `#1'. Using the pattern for}%
\typeout{** the default language instead.}%
\else
\language=\csname l@#1\endcsname
\fi
#2}}
\providecommand{\BIBdecl}{\relax}
\BIBdecl

\bibitem{stevens2020ai}
R.~Stevens \emph{et~al.}, ``{AI} for science,'' Argonne National Lab.(ANL),
  Argonne, IL (United States), Tech. Rep., 2020, report number: ANL-20/17.

\bibitem{shumate2018iot}
J.~Shumate \emph{et~al.}, ``{IoT} for real-time measurement of high-throughput
  liquid dispensing in laboratory environments,'' \emph{SLAS Technology},
  vol.~23, no.~5, pp. 440--447, 2018.

\bibitem{mourtzis2016industrial}
D.~Mourtzis \emph{et~al.}, ``Industrial big data as a result of iot adoption in
  manufacturing,'' \emph{Procedia cirp}, vol.~55, pp. 290--295, 2016.

\bibitem{en14123620}
\BIBentryALTinterwordspacing
A.~D. Rocha \emph{et~al.}, ``Event-driven interoperable manufacturing ecosystem
  for energy consumption monitoring,'' \emph{Energies}, vol.~14, no.~12, 2021.
  [Online]. Available: \url{https://www.mdpi.com/1996-1073/14/12/3620}
\BIBentrySTDinterwordspacing

\bibitem{xu2018industry}
L.~D. Xu \emph{et~al.}, ``Industry 4.0: state of the art and future trends,''
  \emph{International Journal of Production Research}, vol.~56, no.~8, pp.
  2941--2962, 2018.

\bibitem{mourtzis2016cloud}
D.~Mourtzis \emph{et~al.}, ``Cloud-based adaptive process planning considering
  availability and capabilities of machine tools,'' \emph{Journal of
  Manufacturing Systems}, vol.~39, pp. 1--8, 2016.

\bibitem{kanawaday2017machine}
A.~Kanawaday and A.~Sane, ``Machine learning for predictive maintenance of
  industrial machines using iot sensor data,'' in \emph{2017 8th IEEE
  International Conference on Software Engineering and Service Science
  (ICSESS)}.\hskip 1em plus 0.5em minus 0.4em\relax IEEE, 2017, pp. 87--90.

\bibitem{marjani2017big}
M.~Marjani \emph{et~al.}, ``Big {IoT} data analytics: Architecture,
  opportunities, and open research challenges,'' \emph{IEEE Access}, vol.~5,
  pp. 5247--5261, 2017.

\bibitem{pan2021flame}
J.~Pan \emph{et~al.}, ``Flame stability analysis of flame spray pyrolysis by
  artificial intelligence,'' \emph{The International Journal of Advanced
  Manufacturing Technology}, vol. 114, no.~7, pp. 2215--2228, 2021.

\bibitem{xue2019deepfusion}
H.~Xue \emph{et~al.}, ``Deepfusion: A deep learning framework for the fusion of
  heterogeneous sensory data,'' in \emph{Proceedings of the Twentieth ACM
  International Symposium on Mobile Ad Hoc Networking and Computing}, 2019, pp.
  151--160.

\bibitem{dlhub}
R.~Chard \emph{et~al.}, ``{DLHub}: Model and data serving for science,'' in
  \emph{Intl Parallel and Distributed Processing Symp.}\hskip 1em plus 0.5em
  minus 0.4em\relax IEEE, 2019, pp. 283--292.

\bibitem{liu2021bridge}
Z.~Liu \emph{et~al.}, ``Bridge data center ai systems with edge computing for
  actionable information retrieval,'' in \emph{The 3rd Annual Workshop on
  Extreme-Scale Experiment-in-the-Loop Computing}, 2021.

\bibitem{elias2019mdml}
J.~Elias \emph{et~al.}, ``The manufacturing data and machine learning platform:
  Enabling real-time monitoring and control of scientific experiments via
  iot,'' 2020.

\bibitem{SOKOLOWSKI1977219}
\BIBentryALTinterwordspacing
M.~Sokolowski \emph{et~al.}, ``The “in-flame-reaction” method for al2o3
  aerosol formation,'' \emph{Journal of Aerosol Science}, vol.~8, no.~4, pp.
  219 -- 230, 1977. [Online]. Available:
  \url{http://www.sciencedirect.com/science/article/pii/0021850277900416}
\BIBentrySTDinterwordspacing

\bibitem{parkhill1966challenge}
D.~Parkhill, \emph{The Challenge of the Computer Utility}.\hskip 1em plus 0.5em
  minus 0.4em\relax Addison-Wesley, 1966.

\bibitem{foster2001anatomy}
I.~Foster \emph{et~al.}, ``The anatomy of the grid: Enabling scalable virtual
  organizations,'' \emph{The International Journal of High Performance
  Computing Applications}, vol.~15, no.~3, pp. 200--222, 2001.

\bibitem{milojicic2002peer}
D.~S. Milojicic \emph{et~al.}, ``Peer-to-peer computing,'' 2002.

\bibitem{vescovi2022linking}
R.~Vescovi \emph{et~al.}, ``Linking instruments and {HPC}: Patterns,
  technologies, experiences,'' 2022, arxiv.

\bibitem{kreps2011kafka}
J.~Kreps \emph{et~al.}, ``Kafka: A distributed messaging system for log
  processing,'' in \emph{Proceedings of the NetDB}, vol.~11, 2011, pp. 1--7.

\bibitem{funcx}
R.~Chard \emph{et~al.}, ``Funcx: A federated function serving fabric for
  science,'' in \emph{Proceedings of the 29th International Symposium on
  High-Performance Parallel and Distributed Computing}, 2020, pp. 65--76.

\bibitem{chard2016globus}
K.~Chard \emph{et~al.}, ``Globus: Recent enhancements and future plans,'' in
  \emph{XSEDE Conference}.\hskip 1em plus 0.5em minus 0.4em\relax ACM, 2016,
  p.~27.

\bibitem{argonne-discovery-cloud}
\BIBentryALTinterwordspacing
 [Online]. Available: \url{https://github.com/AD-SDL/adc-rdm-sdk}
\BIBentrySTDinterwordspacing

\bibitem{merf-fsp-paper}
M.~Levental \emph{et~al.}, ``Towards online steering of flame spray pyrolysis
  nanoparticle synthesis,'' 2020.

\bibitem{NEURIPS2019_9015}
\BIBentryALTinterwordspacing
A.~Paszke \emph{et~al.}, ``Pytorch: An imperative style, high-performance deep
  learning library,'' in \emph{Advances in Neural Information Processing
  Systems 32}, H.~Wallach \emph{et~al.}, Eds.\hskip 1em plus 0.5em minus
  0.4em\relax Curran Associates, Inc., 2019, pp. 8024--8035. [Online].
  Available:
  \url{http://papers.neurips.cc/paper/9015-pytorch-an-imperative-style-high-performance-deep-learning-library.pdf}
\BIBentrySTDinterwordspacing

\bibitem{merf-fsp-github}
\BIBentryALTinterwordspacing
(May 19, 2020). [Online]. Available:
  \url{https://github.com/globus-labs/MERF-FSP}
\BIBentrySTDinterwordspacing

\bibitem{mcinnes2018umap}
L.~McInnes \emph{et~al.}, ``Umap: Uniform manifold approximation and projection
  for dimension reduction,'' \emph{arXiv preprint arXiv:1802.03426}, 2018.

\bibitem{hdscan}
R.~J. G.~B. Campello \emph{et~al.}, ``Density-based clustering based on
  hierarchical density estimates,'' in \emph{Advances in Knowledge Discovery
  and Data Mining}, J.~Pei \emph{et~al.}, Eds.\hskip 1em plus 0.5em minus
  0.4em\relax Berlin, Heidelberg: Springer Berlin Heidelberg, 2013, pp.
  160--172.

\bibitem{edbo-paper}
B.~Shields \emph{et~al.}, ``Bayesian reaction optimization as a tool for
  chemical synthesis,'' \emph{Nature}, vol. 590, pp. 89--96, 02 2021.

\bibitem{jonas2019rapid}
E.~Jonas and S.~Kuhn, ``Rapid prediction of nmr spectral properties with
  quantified uncertainty,'' \emph{Journal of cheminformatics}, vol.~11, no.~1,
  pp. 1--7, 2019.

\bibitem{parmoo}
\BIBentryALTinterwordspacing
 [Online]. Available: \url{https://github.com/parmoo}
\BIBentrySTDinterwordspacing

\bibitem{doi:10.1080/23307706.2014.885288}
\BIBentryALTinterwordspacing
C.~G. Cassandras, ``The event-driven paradigm for control, communication and
  optimization,'' \emph{Journal of Control and Decision}, vol.~1, no.~1, pp.
  3--17, 2014. [Online]. Available:
  \url{https://doi.org/10.1080/23307706.2014.885288}
\BIBentrySTDinterwordspacing

\bibitem{10.1145/3373464.3373470}
\BIBentryALTinterwordspacing
H.~M. Gomes \emph{et~al.}, ``Machine learning for streaming data: State of the
  art, challenges, and opportunities,'' \emph{SIGKDD Explor. Newsl.}, vol.~21,
  no.~2, p. 6–22, nov 2019. [Online]. Available:
  \url{https://doi.org/10.1145/3373464.3373470}
\BIBentrySTDinterwordspacing

\bibitem{Happ2017MeetingIP}
D.~Happ \emph{et~al.}, ``Meeting iot platform requirements with open pub/sub
  solutions,'' \emph{Annals of Telecommunications}, vol.~72, pp. 41--52, 2017.

\bibitem{SAHAL2020138}
\BIBentryALTinterwordspacing
R.~Sahal \emph{et~al.}, ``Big data and stream processing platforms for industry
  4.0 requirements mapping for a predictive maintenance use case,''
  \emph{Journal of Manufacturing Systems}, vol.~54, pp. 138--151, 2020.
  [Online]. Available:
  \url{https://www.sciencedirect.com/science/article/pii/S0278612519300937}
\BIBentrySTDinterwordspacing

\end{thebibliography}

\end{document}